\documentstyle[12pt]{article}
\begin{document}
\begin{center}
{\large\bf ON DISPERSIONLESS HIROTA\\TYPE EQUATIONS\\}
\vspace{25pt}
{\bf Robert Carroll\\}
{\it Mathematics Department\\University of Illinois\\
Urbana, IL 61801\\}
\vspace{25pt}
{\bf Abstract\\}
\end{center}
\vspace{10pt}
\hspace{.5cm} Various connections between 2-D gravity and KdV, dKdV, inverse
scattering,\\
\indent etc. are established.  For KP we show how to extract
from the dispersionless\\
\indent limit of the Fay differential identity of
Takasaki-Takebe the collection of diff-\\
\indent erential equations for $F =
log(\tau^{dKP})$ which play the role of Hirota type equa-\\
\indent tions in the dispersionless theory.
\\[.6cm]\noindent {\bf 1. HIROTA EQUATIONS}
\\[.4cm]\indent In [7] we showed how second derivatives of $F = log
(\tau^{dKP})$ survive in the dispersionless limit of the KP Hirota
equations and in [31] it was shown how such second derivatives are
contained in the dispersionless limit of the Fay differential identity
(cf. [1]).  We show here how to extract all such dispersionless Hirota
type equations from the limiting Fay identity and indicate the pattern.
Thus, referring to [1,6-9,31,32] for background in KP theory,
we recall first some basic relations involving the Lax and gauge
operator. Thus
\medskip
$$ L = \partial + \sum_1^{\infty}u_{n+1}\partial^{-n};\,\,
W = 1 + \sum_1^{\infty}w_n\partial^{-n}
\eqno(1.1)$$
\medskip
\noindent One has $L = W\partial W^{-1}$ and $\partial_nW W^{-1} =
-L_{-}^n$ (Sato equation).  Then define $\psi = W exp(\xi) = w(x,\lambda)
exp(\xi)\,\,(w(x,\lambda) = 1 + \sum_1^{\infty}w_n\lambda^{-n})$ and
$\psi^{*} = W^{*-1}exp(-\xi) = w^{*}(x,\lambda)exp(-\xi)\,\,
(w^{*}(x,\lambda) = 1 + \sum_1^{\infty}w^{*}_j\lambda^{-j})$.  It
follows that
\medskip
$$ B_n = L^n_{+};\,\, \xi = \sum_1^{\infty}x_n\lambda^n\,\,
(x = x_1);\,\, L\psi = \lambda\psi;\,\, \partial_n\psi = B_n\psi\,\,
(\partial_n = \partial/\partial x_n);\eqno(1.2)$$
$$L^{*}\psi^{*} = \lambda\psi^{*};\,\,\partial_n\psi^{*} = -B_n^{*}\psi^{*};
\,\, \partial_nL = [B_n,L];\,\, \partial_nB_m - \partial_mB_n = [B_n,B_m]$$
\medskip
\noindent One also has a tau function $\tau(x)$ and vertex operators
$X, X^{*}$ which satisfy
$$ \psi(x,\lambda) = X(\lambda)\tau/\tau = e^{\xi}\tau_{-}/\tau =
e^{\xi}G_{-}(\lambda)\tau/\tau = e^{\xi}\tau(x_j - 1/j\lambda^j)/\tau;\,\,
G_{\pm}(\lambda) = \eqno(1.3)$$
$$exp(\pm \xi(\tilde{\partial},\lambda^{-1}));\,\,\tilde{\partial} =
(\partial_1,\frac{1}{2}\partial_2,...);\,\, \psi^{*} = X^{*}(\lambda)
\tau/\tau = e^{-\xi}G_{+}(\lambda)\tau/\tau = e^{-\xi}\tau_{+}/\tau$$
\medskip
\noindent One writes also $exp(\xi) = exp(\sum_1^{\infty}x_n\lambda^n) =
\sum_0^{\infty}p_j\lambda^j$ (Schur polynomials).
\\[3mm]\indent For dispersionless KP one takes $\epsilon \to \epsilon x
= X$ and $t_n \to \epsilon t_n = T_n$ in say the KP equation $u_t =
K(u) = u''' + 3uu' + \frac{3}{4}\partial^{-1}\partial_2^2 u$ with
$\partial_n \to \epsilon \partial/\partial T_n$ and $u(x,t_n) \to
\tilde{u}(X,T_n)$ to obtain $\partial_T \tilde{u} = 3\tilde{u}
\partial_X \tilde{u} + \frac{3}{4}\partial^{-1}\partial^2\tilde{u}/
(\partial T_2)^2$ when $\epsilon \to 0$ (see [10] for a more geometric
approach).  Thus, using $(t_n)$ for
$(x_n),$ consider $L = \epsilon\partial + \sum_1^{\infty}u_{n+1}
(\epsilon,T)(\epsilon \partial)^{-n}$.  We think of replacing $(t_n)$
by $(T_n)$ now and assuming $u_n(\epsilon,T) = u_n(T) + O(\epsilon)$
etc. (in an obvious abuse of notation which seems to cause no confusion).
Take then
\medskip
$$ \psi = (1 + O(1/\lambda))e^{(\sum_1^{\infty}T_n\lambda^n/\epsilon)}
= e^{(\frac{1}{\epsilon}S(T,\lambda) + O(1))};
\tau = e^{(\frac{1}{\epsilon^2}F(T) + O(\frac{1}{\epsilon}))}\eqno(1.4)$$
\medskip
\noindent and replace $\partial_n$ by $\epsilon\partial_n$ where
$\partial_n \sim \partial/\partial T_n$ now to get $L\psi = \lambda\psi$ with
$$ \partial_n L = [B_n,L];\,\,
\epsilon\partial_n\psi = B_n\psi;\,\,\psi =
e^{(\sum_1^{\infty}\frac{1}{\epsilon}
T_n\lambda^n)}\tau(\epsilon,T_n - \epsilon/n\lambda^n)/
\tau(\epsilon,T) \eqno(1.5)$$
\medskip
\noindent The connection to standard semiclassical approximation ideas is
clear.  Now observe that for $P = \partial S\,\,(\partial \sim \partial/
\partial X$ now), $\epsilon^i\partial^i\psi \to P^i \psi$ as $\epsilon
\to 0$ since $\epsilon\partial\psi = \partial S\psi = P\psi,\,\,
\epsilon^2\partial^2\psi = (\epsilon\partial P)\psi + P^2\psi$, etc.
Further $\epsilon\partial(\psi/P) = -((\epsilon\partial P)/P^2)\psi
+ \psi = \psi + O(\epsilon)\psi$ which implies $(\epsilon\partial^{-1})
\psi \to P^{-1}\psi$ etc.  Hence $L\psi = \lambda\psi$ becomes
$$ \lambda = P + \sum_1^{\infty}u_{n+1}P^{-n};\,\,P = \lambda -
\sum_1^{\infty}P_i\lambda^{-i} \eqno(1.6)$$
\medskip
\noindent (we write $u_{n+1}(T)$ or $u_{n+1}(t)$ depending on context).
Also from $B_n\psi = \sum_0^nb_{mn}(\epsilon\partial)^m\psi$ one gets
$$ B_n\psi = \sum_0^nb_{mn}P^m\psi + O(\epsilon)\psi = {\cal B}_n(P)\psi
+ O(\epsilon)\psi;\,\,\epsilon\partial_n\psi = \partial_n S\psi =
{\cal B}_n(P)\psi \eqno(1.7)$$
$$+ O(\epsilon)\psi \Rightarrow \partial_n S = {\cal B}_n(P)
\Rightarrow \partial_n P = \hat{\partial}{\cal B}_n(P)\,\, (\hat{\partial}
{\cal B}_n = \partial_X {\cal B}_n + \partial_P{\cal B}_n(\partial P/
\partial X))$$
\medskip
\noindent We write $B_n = L^n_{+} \to {\cal B}_n = \lambda^n_{+}$
(where + referes to powers of P now) and we note that $(L,\lambda)
\to (\lambda,P)$ (the notation is different in [31,32] for example
where in particular k is used for P).  One has a conventional notation
for $A = \sum a_i\partial^i \sim \sum a_i\xi^i,$ etc., namely $[A,B]
= A \circ B - B \circ A;\,\,A \circ B = \sum_0^{\infty}\partial^n_{\xi}
A\partial^n B/n! = A(x,\xi)exp(\overleftarrow{\partial_{\xi}}
\overrightarrow{\partial_X})B(x,\xi)$.  Putting $\xi = \epsilon\partial$
the term of first order in $\epsilon$ is $\{A,B\} = \partial_{\xi}A
\partial B - \partial A\partial_{\xi}B = \sum \sum (ia_i b_{jX}
-jb_j a_{iX})\xi^{i+j+1}$.  Thinking of action on $\psi$ one replaces
$\xi$ by P and obtains (${\cal B} \sim \lambda^n_{+}):\,\,\partial_n
\lambda = \{{\cal B}_n,\lambda\}.$  Further, putting in $\epsilon$ at
appropriate places one obtains from [31,32] ($S = S(T_n,\lambda),\,\,
T_1 = X,\,\,S_j \to S_j(T_n)$)
$$ \partial_n{\cal B}_m - \partial_m{\cal B}_n + \{{\cal B}_m,{\cal B}_n\}
= 0;\,\, \partial_n S = {\cal B}_n;\,\,S = \sum_1^{\infty}T_n\lambda^n
+ \sum_1^{\infty}S_{i+1}\lambda^{-i} \eqno(1.8)$$
\medskip
\noindent and we note explicitly ${\cal B}_1 = \lambda_{+} = P$.
Further ${\cal B}_n = \lambda^n + \sum_1^{\infty}\partial_n S_{i+1}
\lambda^{-i}$ and one shows that $\partial S_{j+1} = -P_j$. Finally
the tau function for dKP is simply defined via $\partial_n log(\tau^{dKP}) =
-nS_{n+1}\,\,(n\geq 1)$ where $log(\tau^{dKP}) \sim F$ and hence
$\partial S_{n+1} \sim -P_n \sim \partial \partial_n F/n$.
\\[3mm]\indent Now the Hirota bilinear identity for KP is $\oint_{\infty}
\psi(x,\lambda)\psi^{*}(y,\lambda)d\lambda = 0$ and this contains all
of the Hirota bilinear equations for KP (cf. [6,7]).  Such equations
correspond to Plucker relations and in fact a subset of such relations
embodied in the Fay differential identity is also equivalent to the KP
hierarchy (cf. [31]).  The Fay identity can be written (cf. [1] -
$[s_i] = (s_i,\frac{1}{2}s_i^2,\frac{1}{3}s_i^3,...)$ and c.p. $\sim$
cyclic permutations of $1,2,3$)
$$\sum_{c.p.}(s_0 - s_1)(s_2 - s_3)\tau(t + [s_0] + [s_1])\tau
(t + [s_2] + [s_3]) = 0 \eqno(1.9)$$
\medskip
\noindent Differentiating this in $s_0$ and manipulating variables
suitably leads to the differential Fay identity (with $\epsilon$
included for dKP limits - cf. [1,31])
$$ \epsilon\partial_X\tau(t - \epsilon [\mu^{-1}])\tau(t - \epsilon
[\lambda^{-1}]) - \tau(t - \epsilon [\mu^{-1}])\epsilon\partial_X\tau(t -
\epsilon [\lambda^{-1}]) \eqno(1.10)$$
$$- (\lambda - \mu)\tau(t - \epsilon [\mu^{-1}])\tau(t - \epsilon
[\lambda^{-1}]) + (\lambda - \mu)\tau(t)\tau(t - \epsilon [\mu^{-1}] -
\epsilon [\lambda^{-1}]) = 0$$
\medskip
\noindent($X \sim T_1$).  Then in [31] it is shown that this has a
dispersionless limit ($\epsilon \to 0, F = log(\tau^{dKP}))$
$$ \sum_{m,n=1}^{\infty} \mu^{-m}\lambda^{-n}\frac{\partial_m
\partial_n F}{mn} = log[1 + \sum_1^{\infty}\frac{\lambda^{-n} -
\mu^{-n}}{\mu - \lambda}\frac{\partial\partial_n F}{n}] \eqno(1.11)$$
\medskip
\noindent Hence the partial differential equations for F arising from
(1.11) should characterize the dKP system.  Now, although (1.11)
provides a collection of partial differential equations for F in principle,
their determination is not immediate and we provide here an effective
way of determining these equations.  This rests upon an observation
made in [10].  Thus in [10] we used (1.6) $(P = \lambda -
\sum_1^{\infty}P_i\lambda^{-i}$) and $\partial \partial_n F = nP_n$
to write
$$ \sum_1^{\infty}\lambda^{-n}\frac{\partial \partial_n F}{n} =
\sum_1^{\infty}\lambda^{-n}P_n = \lambda - P(\lambda) \eqno (1.12)$$
\medskip
\noindent Hence the right side of (1.11) becomes $log(\frac{P(\mu) -
P(\lambda)}{\mu - \lambda})$.  Thus for $\mu \to \lambda$ and
$\dot{P} \sim \partial_{\lambda}P$
$$ log\dot{P}(\lambda) = \sum_{m,n=1}^{\infty}\lambda^{-m-n}
\frac{\partial_m\partial_n F}{mn} \eqno(1.13)$$
\medskip
\noindent But we already know via (1.6) (since $P_n = \partial
\partial_n F/n$)
$$ \dot{P}(\lambda) = 1 + \sum_1^{\infty} jP_j\lambda^{-j-1} = 1 +
\sum_1^{\infty}\partial \partial_j F\lambda^{-j-1} \eqno(1.14)$$
\medskip
\noindent Looking at the first terms in the expansion of $log(\dot{P}
(\lambda))$ from (1.14) one has ($log(1+x) = x - \frac{1}{2}x^2 +
\frac{1}{3}x^3 + ...$)
$$ log(\dot{P}) \sim \partial^2F\lambda^{-2} + \partial\partial_2 F
\lambda^{-3} + \partial\partial_3 F\lambda^{-4} + \partial\partial_4 F
\lambda^{-5} + ... \eqno(1.15)$$
$$ - \frac{1}{2}[(\partial^2 F)^2\lambda^{-4} + 2\partial^2 F
\partial\partial_2 F\lambda^{-5} + ...] + \frac{1}{3}[(\partial^2 F)^3
\lambda^{-6} + ...] + ...$$
\medskip
\noindent Hence the consistency equations with (1.13) begin with
$\partial^2 F = \partial^2 F$, and $\partial \partial_2 F = \partial
\partial_2 F$, which are tautological, and thereafter we obtain
in particular
$$ \frac{1}{2}(\partial^2 F)^2 - \frac{1}{3}\partial\partial_3 F +
\frac{1}{4}(\partial_2^2 F) = 0;\,\, \partial^2 F \partial\partial_2 F -
\frac{1}{2}\partial\partial_4 F + \frac{1}{3}\partial_2\partial_3 F = 0;
\eqno(1.16)$$
$$-\frac{1}{3}(\partial^2 F)^3 + \partial^2 F\partial\partial_3 F + \frac{1}{2}
(\partial\partial_2 F)^2 -\frac{3}{5}\partial\partial_5 F + \frac{1}{9}
\partial_3^2 F + \frac{1}{4}\partial_2\partial_4 F = 0;...$$
\\[3mm]\indent {\bf THEOREM 1.1}$\,\,$  The relations (1.16) obtained from
(1.13)-(1.14) are a system of partial differential equations for $F =
log(\tau^{dKP})$ which should characterize F and thus play the role
of dispersionless Hirota equations.
\\[3mm]\indent {\bf REMARK 1.2}$\,\,$  It might be interesting to utilize the
equations (1.16) in the study of the WDVV equations of [17] which
classify certain 2-D topological field theories. Mr.
Seung Hwan Son has shown how the equations (1.16) can be generated via a
Mathematica program.  We note also that the
quantity $P(\lambda) - P(\mu)$ is an important ingredient in the
Hamilton-Jacobi theory of Kodama-Gibbons (cf. [7,8,19,29]).
\\[.6cm]\noindent {\bf 2. KdV, mKdV, dKdV and GRAVITY}.
\\[.4cm]\indent We indicate here a few observations which will be
greatly expanded in [11].  First, referring to [7,8] one recalls that
a dispersionless Jevicki-Yoneya action principle was derived for the
KdV-Hermitian matrix model (HMM) situation based on an action (cf.
[7,8,23,34])
$$ \tilde{{\cal A}} = \int Res_P(-\hat{{\cal P}}{\cal Q} - S + \xi)dX
\eqno(2.1)$$
\medskip
\noindent and in [10] it was shown that in fact the term $S - \xi$
arises as a dispersionless version of the Sato equation $\partial_n W
W^{-1} = -L^n_{-}$ since for $W \sim exp(-\frac{1}{\epsilon}H),\,\,
-H \sim S - \xi$, and
$\epsilon\partial_n W W^{-1} \to \partial_n(S - \xi)$ so $\partial_n
(S - \xi) = -\lambda^n_{-}$.
It could be interesting to examine further the meaning here
of this dispersionless Sato equation as an action term.  We note also
that e.g. the conformal flows pass to a dispersionless form via Sato
type equations.  Thus (cf. [1,9,16,21,26]) one defines an Orlov
operator M via
$$ \partial_{\lambda}\psi = M\psi = W(\sum_1^{\infty}kx_k\partial^{k-1})
W^{-1}\psi = (G + \sum_2^{\infty}kx_k\lambda^{k-1})\psi; G = WxW^{-1}
\eqno(2.2)$$
\medskip
\noindent and the conformal flows are $\partial_{m+1,1}u = \partial
Res(ML^{m+1})$ with $\partial_{m+1,1}W = -(ML^{m+1})_{-}W$.  One knows
(cf. [7,8,31,32]) that $M \to {\cal M} = S_{\lambda}$ and $\epsilon
\partial_{m+1,1}W_{\epsilon}W_{\epsilon}^{-1} \to -({\cal M}
\lambda^{m+1})_{-}$ leading to $\partial_{m+1,1}(S - \xi) =
-({\cal M}\lambda^{m+1})_{-}$
with similar results for other additional symmetry flows.
In this context we recall also that $(L,G) \to (\lambda,\hat{\xi})$
where $(\lambda,-\hat{\xi})$ are action angle variables in the
Hamilton-Jacobi theory of [19,29].
\\[3mm]\indent Now based on a less sophisticated action idea for HMM
situations (cf. [7,8,20]) $(\delta/\delta u)\int R_{k+1}(u)dx =
-(k+\frac{1}{2})R_k(u)$, where $R_k \sim$ Gelfand-Dickey resolvant terms,
one develops a unitary matrix model (UMM) action via (cf. [2-4,14,15])
$$ I = \int \sum_1^{\infty}[4t_{2k+1}R_{k+1} - v^2x]dx \eqno(2.3)$$
\medskip
\noindent where $u = v^2 - v', \delta u = 2v\delta v - \partial\delta v,
\hat{D} = \partial + 2v$, so that $(\delta I/\delta v) = -\sum_1^{\infty}
(2k+1)t_{2k+1}\hat{D}R_k - vx = 0$.  Since $R_k \to \frac{1}{2}r_{k-1}(p)\,\,
(p = -\frac{1}{2}u)$ in the dispersionless limit (cf. [7,8]) with
$r_{n-1}(p) = (n-\frac{1}{2})...\frac{1}{2}/n!(2p)^n$ and $\epsilon
\partial R_k \to 0$, upon shifting to $T'$ variables $(\epsilon t_n =
T_n,\,nT_n = T'_n$), the UMM yields a limiting string equation
$$ \sum_1^{\infty}(2k+1)T'_{2k+1}r_{k-1}(p) = -X \eqno(2.4)$$
\medskip
\noindent which is in fact the Landau-Ginsburg (LG) equation of HMM. Thus
\\[3mm]\indent {\bf PROPOSITION 2.1}$\,\,$ In the dispersionless limit
the UMM, based on mKdV, corresponds to the HMM, based on KdV, via the
LG equation.
\\[3mm]\indent This is meant to be heuristic since there has been little
physical input; one argues in in [15] for example that KdV and mKdV are
descriptions of an open-closed string theory involving gravity.  In
particular, the unique real nonsingular solutions provided by UMM
map to open string generalizations of the nonsingular solutions of the
KdV hierarchy found in [14].  Those solutions in turn arise by first
assuming that the KdV flows are preserved nonperturbatively, which leads
uniquely to the HMM string equations for closed strings, and to a unique
pole free string susceptibility $\sim u$.  For such closed (resp. open)
strings one is dealing with HMM and Painleve 1 (resp. UMM and Painleve 2).
The details here are complicated and we will say more about this in [11].
Further gravity topics in [11] will include material from [25] on a direct
passage from KdV to gravity and from [22,27-30,33] on Liouville-
Beltrami gravity and connections to KdV, mKdV, and Ur-KdV.
\\[.6cm]\noindent {\bf 3. ON INVERSE SCATTERING AND DISPERSIONLESS
LIMITS}
\\[.4cm]\indent In [7,8] we outlined some connections between inverse
scattering for KdV and mKdV, dKdV.  This is developed in more detail
and generality in [11,13] (cf. also [5]).  Thus in dKdV with $\epsilon
x = X,\,\,x \to \pm\infty \sim \epsilon \to 0$, so there should be a
natural relation between some inverse scattering quantities and some
corresponding dispersionless quantities.  Here the analysis of [18,19,35]
is particularly relevant.  We recall that $u_t = u''' - 6uu', u = v^2 - v',
v_t = v''' - 6v^2v'$ are related and via a Galilean transformation
$u \to u - \lambda, t \to t, x \to x - 6\lambda t$, KdV is unaltered while
$mKdV \to v_t = 6\lambda v' + v''' -6v^2v'$.  Then for $u + \lambda =
v^2 - v'$ with $v = -\psi'/\psi$ one has $\psi'' - u\psi = \lambda\psi
= -k^2\psi$ with Jost solutions $\psi_{\pm}(k,x) \sim exp(\pm ikx)$ as
$x \to \pm\infty$.  We recall that the transmission coefficient T and
reflection coefficients $R,R_L$ arise in formulas $T(k)\psi_{-} =
R(k)\psi_{+} + \psi_{+}(-k,x);\,\,T(k)\psi_{+} = R_L(k)\psi_{-} +
\psi_{-}(-k,x)$ and we set $\psi_{-} = exp(-ikx + \phi(k,x))$ with
$\phi(k,-\infty) = 0$.  This leads to (*) $\phi'' - 2ik\phi' +
\phi'^2 = u$, so $-\psi'_{-}/\psi_{-} = -(-ik + \phi') = v$ and one
expects a solution of (*) in the form $\phi' = \sum_1^{\infty}
\phi_n/(ik)^n$ compatible with $u + \lambda = v^2 - v'$, i.e. $v \sim
ik + \sum_1^{\infty}v_n/(ik)^n$.  This leads to $\phi_n = v_n$.
\\[3mm]\indent Next recall (cf. [6]) that for suitable u with say
$\int_{-\infty}^{\infty}(1 + |x|^2)|u|dx < \infty$ one has available
all of the inverse scattering machinery and for convenience one could
even assume $u \in {\cal S}$ (Schwartz space).  One expects T(k) to be
meromorphic for $Im(k) > 0$ with (possibly) a finite number of simple
poles at $\beta_j = ik_j$.  Further (cf. [6]) $1/T = (1/2ik)W(\psi_{+},
\psi_{-}) = (1/2ik)(\psi'_{+}\psi_{-} - \psi_{+}\psi'_{-})$.  Given
$|R(k)|$ small for large $|k|$ real (which will be true for $u \in {\cal S}$
say) one expands $1/(\xi - k) = -\sum_0^{\infty}(\xi^n/k^{n+1})$ and writes
($\bar{R}(k) = R(-k)$ for k real)
$$ log(T) \sim \sum_0^{\infty}\frac{c_{2n+1}}{k^{2n+1}}; c_{2n+1} =
\frac{2}{2n+1}\sum_1^N(i\beta_j)^{2n+1} - \frac{1}{2\pi i}
\int_{-\infty}^{\infty}\xi^{2n}log(1 - |R|^2)d\xi \eqno(3.1)$$
\medskip
\noindent Now for $x \to \infty,\,\,\psi_{-}e^{ikx} \sim (R/T)e^{2ikx}
+ (1/T) \to 1/T$ when $Im(k) > 0$.  Taking logarithms we get $log(\psi_{-})
+ ikx = \phi(k,x) \to -log(T)$ as $x \to \infty$ so $\phi(k,\infty) =
-log(T)$.  Hence $\int_{-\infty}^{\infty}\phi_{2m}dx = 0$ and
$$ -log(T) = \sum_1^{\infty}\frac{1}{(ik)^n}\int_{-\infty}^{\infty}
\phi_ndx = -\sum_0^{\infty}\frac{c_{2n+1}}{k^{2n+1}};\,\,i^{2m}c_{2m+1} =
\int_{-\infty}^{\infty}\phi_{2m+1}dx \eqno(3.2)$$
\\[3mm]\indent {\bf THEOREM 3.1}.$\,\,$ The scattering information or
x-asymptotic information, given via R,T is related to the
k or $\lambda$ asymptotics of the wave function $\psi_{-}$, given via
$\phi_n$, through the formulas (3.2), and $\phi_n = v_n$ establishes a
further connection to the mKdV expansion terms.
\\[3mm]\indent We remark that there are arguments based on a Gardner
type transform $u = \alpha\epsilon^2v^2 + v + \beta\epsilon v'$ for
example which lead to an identification of the $c_{2n+1}(u)$ with
Hamiltonians $H_n$.  This tells us that the $\phi_{2n+1}$ are nontrivial
conserved densities and hence so are the corresponding $v_n$.  Now from
the dispersionless theory of Sections 1,2 we have (cf. also [7,8,32])
$-u = q = 2p,\,\,\lambda = -k^2,\,\,W\partial^2 W^{-1} \to \hat{{\cal P}} =
P^2 + q = (ik)^2$, and $ik = P(1 + \sum_1^{\infty}{\frac{1}{2} \choose m}
q^mP^{-2m}$.  This corresponds to (1.6) with $\lambda = ik$ and there will
be a corresponding inversion $P = ik - \sum_1^{\infty}P_j(ik)^{-j}$.
Consider $\psi \sim \psi_{+} = exp(\frac{1}{\epsilon}S)$ with
$\psi^{*} \sim \psi_{-} = exp(-\frac{1}{2}S)$ (cf. [10] for a proof of the
last statement).  We note explicitly that $\epsilon\partial\psi_{-}/\psi_{-}
= - P = -S_X$ and $\epsilon^2\partial^2\psi_{-}/\psi_{-} = P^2 + \epsilon P'
\to
P^2$ leading to $(ik)^2 = P^2 + q$.  But $\psi_{-} = exp(-ikx + \phi)
\sim \psi_{-} = exp(-\frac{1}{\epsilon}S)$ for $|x|$ large or $\epsilon$
small which suggests an identification $-ikX/\epsilon + \phi(X/\epsilon,k)
=-\frac{1}{\epsilon}S(X,k)$ with $-S' = -ik + \phi'$ where $\phi'$ is the
derivative in the x argument.  Here in the same notation where e.g.
$u_n(T,\epsilon) = u_n(T) + O(\epsilon)$ we assume it is permissible to write
$\phi'(X/\epsilon,k) =
\phi'(X,k,\epsilon) = \phi'(X,k) + O(\epsilon)$.  Then one can deduce that
$-ik + \phi' = -S_X = -P; v \sim P$.  Consequently,
\\[3mm]\indent {\bf THEOREM 3.2}$\,\,$  Given the above relations betwen
variables x and X, the dispersionless KdV theory
interacts with scattering data $\phi_n$ and mKdV data $v_n$ via
$\phi_n \sim v_n \sim -P_n$.
\\[2mm]\indent Now consider the use of $cklog|T| = \alpha(k);\,\,
\beta(k) = \hat{c}arg(R_L/T);\,\, Q_n = C_n\beta_n^2;\,\,\hat{P}_n
= \hat{C}_nlog|b_n|$ as action-angle variables in a Hamiltonian theory
for KdV (cf. [6] - the c, $\hat{c},\,C_n,\,\hat{C}_n$ are constants
which depend on notation and the $b_n$ are normalization constants).
Under standard Poisson brackets $\{\alpha(k),\beta(k')\} = \delta(k - k')$
and $\{Q_n,\hat{P}_n\} = \delta_{mn}$ with other brackets vanishing
(note such Gardner type brackets do not go over to our canonical phase
space brackets as in (1.8)).  Now recall that $1/T = (1/2ik)W(\psi_{+},
\psi_{-})$ and consider $\psi_{\pm} \sim exp(\pm \frac{1}{\epsilon}S)$
as above.  Then $W(\psi_{+},\psi_{-}) \sim \epsilon\partial e^{\frac{1}
{\epsilon}S}e^{-\frac{1}{\epsilon}S} - e^{\frac{1}{\epsilon}S}
\epsilon\partial e^{-\frac{1}{\epsilon}S} = 2S' = 2P;\,\,|T| = |k/P|$.
Note here $\epsilon \to 0 \sim x \to \infty$ so there is no contradiction
to W being independent of x while P is just a phase space variable on an
(X,P) phase space.  Hence $\alpha(k) \sim cklog(|k/P|)$.  On the other
hand from [6] $R_L = W(\psi_{-}(-k,x),\psi_{+})/W(\psi_{+},\psi_{-})$
and $\psi_{-}(-k,x) \sim exp(-\frac{1}{\epsilon}S(X,k))$ so the
exponentials do not cancel.  However some analysis based on [18] (details
in [12]) allows us to say
that $Im S = 0$ and $\beta(k) \to \hat{c}[\pi\,\,or\,\,2\pi]$.  Thus
\\[3mm]\indent {\bf THEOREM 3.3}.$\,\,$ The action variable $\alpha(k)
= cklog|T|$ in KdV passes to a phase space variable $cklog|k/P|$ in
dKdV and locally $\beta(k) \to$ constant.

\end{document}